\documentclass[10pt, a4paper]{article}

\usepackage[final]{lrec2026} %

\usepackage{tikz}
\usetikzlibrary{positioning}
\usepackage[T1]{fontenc}
\usepackage{graphicx}
\usepackage{booktabs} %
\usepackage{enumitem}
\usepackage{multirow}
\usepackage{colortbl}
\usepackage[most]{tcolorbox}
\usepackage{array}         %
\usepackage{xcolor}        %
\usepackage{ulem}
\usepackage{makecell} %
\usepackage{amsmath}    %
\newcommand{\code}[1]{{\small\sffamily\itshape #1}}
\usepackage{color}

\title{Do We Need Bigger Models for Science? Task-Aware Retrieval with Small Language Models}

\name{
Florian Kelber$^{1*}$\thanks{* Equal contribution.}, 
Matthias Jobst$^{1*}$\footnotemark[1], 
Yuni Susanti$^{2}$, 
Michael Färber$^{3}$
}

\address{
$^{1}$TU Dresden, Germany \\
$^{2}$FIZ Karlsruhe, Germany \\
$^{3}$ScaDS.AI, TU Dresden, Germany \\
florian.kelber@tu-dresden.de, matthias.jobst2@tu-dresden.de, \\
yuni.susanti@fiz-karlsruhe.de, michael.faerber@tu-dresden.de
}

\abstract{
Scientific knowledge discovery increasingly relies on large language models, yet many existing scholarly assistants depend on proprietary systems with tens or hundreds of billions of parameters. Such reliance limits reproducibility and accessibility for the research community. In this work, we ask a simple question: \textit{do we need bigger models for scientific applications?} Specifically, we investigate to what extent carefully designed retrieval pipelines can compensate for reduced model scale in scientific applications.
We design a lightweight retrieval-augmented framework that performs \textit{task-aware} routing to select specialized retrieval strategies based on the input query.
The system further integrates evidence from full-text scientific papers and structured scholarly metadata, and employs compact instruction-tuned language models to generate responses with citations.
We evaluate the framework across several scholarly tasks, focusing on scholarly question answering (QA), including single- and multi-document scenarios, as well as biomedical QA under domain shift and scientific text compression.
Our findings demonstrate that retrieval and model scale are complementary rather than interchangeable. While retrieval design can partially compensate for smaller models,
model capacity remains important for complex reasoning tasks. This work highlights retrieval and task-aware design as key factors for building practical and reproducible scholarly assistants.
\\ \newline 
}

\begin{document}

\maketitleabstract

\section{Introduction}
\label{sec:intro}

The volume of scientific publications continues to grow rapidly, making it increasingly difficult for researchers to discover and synthesize relevant knowledge. Recent advances in large language models (LLMs) have shown strong potential for supporting scientific tasks such as question answering, summarization, and literature exploration. However, many scholarly assistants rely on proprietary models with tens or hundreds of billions of parameters, creating substantial barriers in terms of computational cost, accessibility, and reproducibility.

Beyond issues of scale, applying general-purpose LLMs to scientific literature presents additional challenges. Scientific documents are highly technical and domain-specific, and models may lack sufficient adaptation to scholarly language and reasoning patterns. As a result, existing systems can produce hallucinated or weakly supported claims when answering scientific questions in research papers~\cite{wadden2024sciriffresourceenhancelanguage,cai2024sciassessbenchmarkingllmproficiency,li2025scilitllmadaptllmsscientific,shen2024tagllmrepurposinggeneralpurposellms}. These limitations raise concerns about reliability and transparency, which are central to scientific applications.

Early efforts to build science-focused language models highlight both the promise and the limitations of current approaches. Systems such as Galactica attempted to encode large amounts of scientific knowledge directly into model parameters, but encountered challenges related to factual accuracy and verification~\cite{taylor2022galactica,galactica2022conversation}. Other approaches incorporate retrieval mechanisms or domain-specific datasets, yet many still depend on large proprietary backbones or remain limited to specific domains. For example, biomedical QA systems built around datasets such as PubMedQA~\cite{jin_pubmedqa_2019} demonstrate the benefits of targeted retrieval, but do not necessarily generalize across the broader scientific ecosystem.

Retrieval-Augmented Generation (RAG) has emerged as a promising paradigm for improving reliability and transparency by grounding model outputs in external evidence~\cite{lewis2021retrievalaugmentedgenerationknowledgeintensivenlp}. In scientific domain, prior work has explored knowledge-graph-based retrieval, domain-specific training, and hybrid approaches that combine multiple sources of evidence~\cite{kg_qa,asaiOpenScholarSynthesizingScientific2024}. 
While these approaches improve grounding, hallucinations remain a persistent challenge~\cite{mishra2024finegrained,mallen-etal-2023-trust}, and many systems continue to rely on large backbone models.

In this work, we examine a simple question: \textit{do we need bigger models for scientific applications?} Specifically, we investigate the extent to which improvements in retrieval design can support the use of smaller models in these settings.
Rather than asking whether retrieval can replace model scaling entirely, we focus on the conditions under which retrieval strategies can support smaller models effectively, and the trade-offs that arise in doing so. In particular, we analyze how retrieval, domain coverage, and task structure interact with model capacity in scholarly question answering and related tasks.

To this end, we design a lightweight retrieval-augmented framework that incorporates \textit{task-aware routing} prior to retrieval. Incoming queries are first classified into scholarly task categories, allowing the system to select specialized retrieval strategies tailored to different information needs. The framework integrates evidence from both unstructured scientific documents and structured scholarly metadata from knowledge graph within a unified pipeline. 
This hybrid design enables support for multiple tasks, including question answering, summarization, and factual metadata queries.

We evaluate the framework across several scholarly tasks, focusing on scholarly QA (ScholarQABench-Multi~\citeplanguageresource{openscholar})--including both single- and multi-document scenarios--as well as biomedical QA under domain shift (PubMedQA~\citeplanguageresource{pubmedqa}) and scientific text compression (SciTLDR~\citeplanguageresource{tldr}).
Our results show that small open-weight models can approach the performance of larger systems when paired with appropriate retrieval strategies. 
However, we also found that system performance remains sensitive to retrieval quality, prompt length, and domain mismatch, highlighting limitations in robustness and generalization. 
Overall, our findings suggest that improved retrieval design can partially compensate for smaller models. Nevertheless, model capacity remains important for complex reasoning tasks, which means that retrieval and model scale are complementary rather than interchangeable. 

Our contributions are summarized as follows:
\begin{itemize}%
\item We design a task-aware routing strategy that selects retrieval methods based on the information needs of scholarly queries.
\item We introduce a hybrid retrieval framework integrating scientific text collections with structured scholarly knowledge graphs.
\item We empirically analyze the extent to which compact open-weight models, combined with targeted retrieval, can approach the performance of larger systems, highlighting key trade-offs in robustness and generalization.
\item We release resources and source code to support reproducibility and further research.\footnote{\url{https://github.com/faerber-lab/lightweight-scholarly-qa}}
\end{itemize}

\section{Related Work}
\label{relwork}

\paragraph{Retrieval-Augmented Systems for Scholarly QA}
Retrieval-Augmented Generation (RAG) has become a dominant paradigm for improving factual grounding in language models~\cite{lewis2021retrievalaugmentedgenerationknowledgeintensivenlp}. In the scholarly domain, several systems combine large language models with retrieval from scientific corpora. For example, OpenScholar retrieves and reranks paper sections from Semantic Scholar, incorporating feedback loops and citation verification mechanisms to improve reliability~\cite{asaiOpenScholarSynthesizingScientific2024}. Large-scale resources such as unarXive~\citeplanguageresource{unarXive} and SciQA~\citeplanguageresource{auerSciQAScientificQuestion2023} provide curated datasets for scientific question answering and retrieval-based experimentation.

While these approaches demonstrate the effectiveness of retrieval in scientific settings, many rely on heavyweight reranking pipelines or large backbone models. In contrast, our work investigates whether a lightweight architecture can achieve competitive performance without complex reranking or large-scale parametric knowledge. Rather than scaling model size, we focus on structuring the retrieval process through task-aware routing.
\vspace{-2mm}
\paragraph{Knowledge Graph-Based Scholarly Information Access}
Knowledge graphs provide a complementary perspective on scholarly knowledge by representing publications, authors, and venues as structured entities and relations. Several large-scale scholarly KGs have been introduced, including SemOpenAlex~\citeplanguageresource{semopenalex}, Semantic Scholar~\citeplanguageresource{kinney2023semanticscholaropendata}, ORKG~\citeplanguageresource{orkg}, and MAKG~\citeplanguageresource{makg}. These resources enable structured querying through languages such as SPARQL and support applications including metadata exploration and scientific discovery. Prior work has investigated machine-learning-driven interfaces that map natural language queries to KG queries. 
Our approach builds on this line of work by integrating KG retrieval within a broader retrieval-augmented generation framework. Rather than focusing exclusively on KG-based QA, we combine structured metadata with text-based evidence to support a wider range of scholarly tasks.

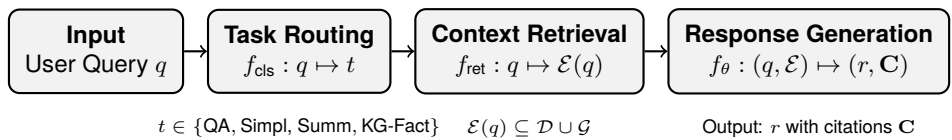
\begin{figure*}[t]
\centering
\begin{tikzpicture}[
    node distance=0.3cm,
    every node/.style={font=\sffamily\footnotesize, align=center},
    box/.style={
        rectangle, rounded corners,
        draw=black, thick,
        minimum width=1.9cm, minimum height=1.1cm,
        text centered,
        fill=gray!10,
        inner sep=6pt
    }
]
\definecolor{lightblue}{RGB}{230,245,255}
\definecolor{lightgreen}{RGB}{230,255,230}
\definecolor{lightorange}{RGB}{255,245,230}
\definecolor{lightpurple}{RGB}{245,230,255}

\node[box] (input) {\textbf{Input}\\User Query $q$};
\node[box, right=of input] (cls) {\textbf{Task Routing}\\$f_{\text{cls}}: q \mapsto t$};
\node[box, right=of cls] (ret) {\textbf{Context Retrieval}\\$f_{\text{ret}}: q \mapsto \mathcal{E}(q)$};
\node[box, right=of ret] (gen) {\textbf{Response Generation}\\$f_\theta: (q,\mathcal{E}) \mapsto (r, \mathbf{C})$};

\draw[->, thick] (input) -- (cls);
\draw[->, thick] (cls) -- (ret);
\draw[->, thick] (ret) -- (gen);

\node[below=0.2cm of cls] {\scriptsize $t \in \{\text{QA}, \text{Simpl, Summ}, \text{KG-Fact}\}$};
\node[below=0.2cm of ret] {\scriptsize $\mathcal{E}(q) \subseteq \mathcal{D} \cup \mathcal{G}$};
\node[below=0.2cm of gen] {\scriptsize Output: $r$ with citations $\mathbf{C}$};

\end{tikzpicture}
\caption{Task-aware retrieval pipeline. Task routing determines which processing strategy and data source are used before response generation. The input query $q$ is classified into a task $t$, used to retrieve relevant context $\mathcal{E}(q)$ from data sources $\mathcal{D}$ and $\mathcal{G}$, and passed to a lightweight LLM to generate the response $r$ with citations $\mathbf{C}$.}
\label{fig:pipeline-flow}
\end{figure*}

\vspace{-2mm}
\paragraph{Plain Language Summarization of Scientific Literature}
Another important line of research focuses on improving the accessibility of scientific documents through plain language summarization (PLS). These systems aim to condense complex texts into clear summaries that can be understood by both expert and non-expert audiences~\cite{know_your_audience_simplification}. Techniques often involve interpreting domain-specific terminology, adding explanatory context, and removing redundant or overly technical details~\cite{guo-etal-2024-appls}. Maintaining factual accuracy while improving readability remains challenging, particularly for specialized domains such as biomedical literature~\cite{joseph-etal-2024-factpico}. Evaluation is also non-trivial: traditional metrics such as ROUGE~\cite{lin-2004-rouge}, BLEU~\cite{bleu}, and BERTScore~\cite{bert-score} capture lexical similarity but often fail to fully measure informativeness or faithfulness~\cite{guo-etal-2024-appls,luo-etal-2022-benchmarking,10.1093/jamia/ocac149}. To evaluate summarization capabilities within our framework, we use the SciTLDR dataset~\cite{cachola-etal-2020-tldr}, which focuses on concise summaries of computer science publications and has been widely used in prior work~\cite{takeshita2024rougeksummarieskeywords}. Earlier experiments using environments such as CATTS extended these tasks with title generation and denoising objectives~\cite{cachola-etal-2020-tldr}.

Despite advances in retrieval and summarization, many scholarly assistants continue to rely on large proprietary models. This dependence limits reproducibility and raises the barrier to entry for academic institutions without extensive computational resources. Recent studies have highlighted both the benefits and limitations of adapting general-purpose LLMs to scientific domains~\cite{wadden2024sciriffresourceenhancelanguage,cai2024sciassessbenchmarkingllmproficiency,li2025scilitllmadaptllmsscientific,shen2024tagllmrepurposinggeneralpurposellms}, as well as the persistence of hallucinations in scholarly settings~\cite{mishra2024finegrained,mallen-etal-2023-trust}.

\paragraph{Positioning of this Work.}
Overall, prior work demonstrates the value of retrieval-augmented generation, domain-specific datasets, and knowledge graphs for scholarly question answering. However, fewer works investigate how these components interact within a lightweight architecture. While prior work often emphasizes scaling model size, fewer studies systematically analyze the trade-offs between retrieval design and model capacity.
Our work addresses these gaps in two key ways. 
First, we explicitly examine whether improved retrieval design can compensate for a reduced model scale. 
Second, we integrate task-aware routing and hybrid retrieval (text + structured scholarly metadata) within a lightweight pipeline. 
We next describe the proposed framework in detail.

\section{Task-Aware Hybrid Retrieval for Scholarly Applications}
\label{sec:pipeline}
Scholarly assistants must handle heterogeneous information needs, ranging from factual metadata queries to multi-document reasoning over scientific literature. Treating these requests uniformly often leads to inefficient retrieval and unnecessary load on large language models. 

We design a task-aware retrieval framework that routes user queries to specialized retrieval strategies before generation. The framework combines (i) task classification and routing with a lightweight classifier, (ii) hybrid context evidence retrieval from both scientific text collections and knowledge graphs, and (iii) response generation using a small language model. Figure~\ref{fig:pipeline-flow} illustrates the overall architecture of the proposed framework.

\vspace{-3mm}
\paragraph{Design Rationale} Our design is guided by two observations. First, many scholarly queries can be solved through targeted retrieval rather than increased model scale. For example, a question ``Which papers propose methods for protein structure prediction?'' can be answered by retrieving relevant abstracts or datasets instead of relying on a massive language model to memorize all literature.  
Second, queries about scholarly metadata are better answered through knowledge graphs or structured database than through text generation alone. For instance, a question ``Who are the co-authors of Zang et al., 2023?'' is more accurately answered by querying structured bibliographic databases or knowledge graphs. These observations motivate the combination of task-aware routing and hybrid retrieval within a lightweight generation framework. 

Formally, given a query $q \in \mathcal{Q}$, the system retrieves evidence
\[
\mathcal{E}(q) = \{e_1, ..., e_M\}
\]
from a textual corpus $\mathcal{D}$ and/or scholarly knowledge graph $\mathcal{G}$. A lightweight language model then produces an answer
\[
r = f_\theta(q, \mathcal{E}(q)).
\]

Each answer is accompanied by a citation set

\[
\mathbf{C} = \{c_1,...,c_K\}, \quad c_i \in \mathcal{D} \cup \mathcal{G},
\]

allowing the system to expose the evidence underlying generated statements. 

\definecolor{rowgray}{gray}{0.96}
\begin{table*}[t]
\centering
\begin{tabular}{>{\raggedright\arraybackslash}m{2.5cm}
                >{\raggedright\arraybackslash}m{6cm}
                >{\raggedright\arraybackslash}m{5.8cm}}
\toprule
\textbf{Task Category} & \textbf{Retrieval Strategy} & \textbf{Output Type} \\
\midrule
\rowcolor{rowgray}

General QA & 
Passage retrieval from unarXive; metadata enrichment via Semantic Scholar API & 
Factual answer with inline citations \\

Simplification, Summarization & 
NER-based paper title detection; fuzzy match against unarXive full-text corpus & 
Simplified or summarized text with optional paper grounding \\

\rowcolor{rowgray}
KG-Fact & 
SPARQL queries over SemOpenAlex; 18 predefined templates (author/work) & 
Structured metadata (e.g., h-index, DOI, ORCID, affiliations) \\

\bottomrule
\end{tabular}
\caption{Overview of task categories, retrieval strategies, and output in the classification module.}
\label{tab:task_summary}
\end{table*}

\subsection{Task Routing}
User queries in scholarly assistants are highly heterogeneous. A request for a paper summary requires different evidence than a factual query asking about an author's affiliation. Applying a single retrieval strategy to all queries therefore introduces unnecessary noise and computational cost. We introduce a lightweight task routing module that predicts the information need prior to retrieval. Given a query $q$, a classifier assigns a task label
\[
f_{\text{cls}} : q \mapsto t \in \mathcal{T}
\]
where
\[
\mathcal{T} =
\begin{aligned}
\{\texttt{General QA},\; \texttt{Simplification},\\
\texttt{Summarization},\; \texttt{KG\text{-}Fact}\}
\end{aligned}
\]

The four task categories reflect distinct information needs that require different retrieval strategies. General QA involves multi-document reasoning over unstructured text, while summarization and simplification require transformation of specific documents. In contrast, KG-Fact queries target structured metadata that can be answered more reliably via knowledge graphs than text retrieval. This taxonomy is not intended to be exhaustive but rather to capture common patterns in scholarly information-seeking. Future work could explore learned or hierarchical task taxonomies.

We implement the classifier using a fine-tuned small language model (\textit{Llama~3.2 3B Instruct}). If classifier confidence falls below a predefined threshold, the system defaults to the general QA retrieval pathway. The predicted task $t$ determines the retrieval strategy used in the next stage, as summarized in Table~\ref{tab:task_summary}.
This routing mechanism reduces unnecessary retrieval noise and allows the system to allocate resources more efficiently.

\subsection{Context Evidence Retrieval}
\label{sec:context_retrieval}
Traditional RAG systems typically rely on unstructured text corpora. However, scholarly information exists both in textual form and in structured metadata sources such as citation graphs and author databases. Unlike conventional RAG pipelines relying solely on unstructured text corpora, our framework integrates both unstructured scientific documents and structured scholarly knowledge graphs. This hybrid design enables the system to answer both narrative questions and precise metadata queries within a unified architecture.

Once a query has been classified into a specific task, the system retrieves task-appropriate \textit{evidence} from curated text corpora or structured knowledge graphs. The context retrieval step thus serves as the bridge between user intent and generation, ensuring that responses are grounded in identifiable sources.
Depending on the task category, one of the following retrieval strategies is invoked, as summarized in Table~\ref{tab:task_summary}.

\paragraph{\textbf{QA: General Scholarly Question Answering.}}
For general QA tasks, text passages are retrieved as evidence from the open subset of the \textit{unarXive} corpus \cite{saierUnarXive2022All2023}, comprising 165K publications. Documents are preprocessed into \textit{markdown} containing metadata and plain text (tables and figures removed). The text is segmented into sections or subsections, further subdivided into chunks of at least 800 characters. Each chunk is embedded using Sentence Transformers (\code{all-MiniLM-L6-v2}) into 384-dimensional vectors. Both vector embeddings and metadata (e.g., \textit{title, authors, venue}) are stored in a FAISS vector index \cite{Douze2024TheFL,Chase_LangChain_2022}.  At query time, the user query $q$ is embedded in the same space, and the top-$k$ most similar chunks based on vector similarity are retrieved:
\[
\mathcal{E}_{\text{QA}}(q) = \text{Top-}k \big(\text{FAISS}( \text{Embed}(q) ) \big).
\]
Retrieved chunks are assigned reference numbers and appended to the user query, following the multi-paper QA prompt design from the OpenScholar \cite{asaiOpenScholarSynthesizingScientific2024}. To enrich bibliographic accuracy, the pipeline also issues secondary lookups to the Semantic Scholar API \cite{kinney2023semanticscholaropendata} using paper titles from the retrieved set. References are tracked for the generation of a corresponding bibliographic list.

\paragraph{\textbf{Simplification and Summarization}} 
For simplification and summarization tasks, the system attempts to detect whether the user refers to a specific scientific paper. A Named Entity Recognizer (NER) component is used to identify candidate paper titles, and if a specific title is detected, a fuzzy match is performed against the corpus (\textit{unarXive} corpus~\citeplanguageresource{unarXive}) to retrieve the full text of the corresponding paper. If no reliable title match is found, the system assumes that the query refers to arbitrary input text rather than a specific paper and performs summarization or simplification directly on the provided input. This fallback ensures that the system remains functional even when grounding is not possible. 

The NER component is implemented using a spaCy \code{EntityRecognizer}\footnote{\url{https://spacy.io/api/entityrecognizer}} trained on synthetic data generated from unarXive titles and manually designed prompt templates. This step enables the system to ground summaries in the full paper when possible, while still supporting general text queries.

\paragraph{\textbf{KG-Fact: Structured Scholarly Metadata Retrieval.}}
Certain scholarly queries request structured metadata such as citation metrics, identifiers, or affiliations. These questions are better addressed through knowledge graphs rather than through text generation alone. For such queries, the system maps user questions to one of 18 predefined SPARQL templates, focusing on author and scientific work-related queries.
A rule-based pre-check detects explicit mentions of scholarly identifiers such as \textit{h-index}, \textit{i10-index}, \textit{ORCID}, or \textit{DOI}. If such identifiers are not present, an NER component extracts candidate author or paper entities, and a classifier assigns the query to the most appropriate SPARQL template. 
The resulting query is executed against the SemOpenAlex knowledge graph via its SPARQL endpoint, returning structured metadata $\mathcal{E}_{\text{KG}}(q)$ in a verifiable format.

Our template-driven approach to knowledge graphs provides a reliable and interpretable interface for knowledge graph access. 
The current implementation uses a fixed set of templates, but these can be readily extended as new query types are added. 

\subsubsection*{\textbf{Final Prompt Composition}}
Across all pathways, context retrieval produces an evidence set $\mathcal{E}(q)$ tailored to the classified task $t$. Formally:
\[
    \mathcal{E}(q, t) = 
    \begin{cases}
        \mathcal{E}_{\text{QA}}(q) & t = \text{\code{QA}}, \\
        \mathcal{E}_{\text{Simpl}}(q) & t = \text{\code{Simpl}}, \\
        \mathcal{E}_{\text{Summ}}(q) & t = \text{\code{Summ}}, \\
        \mathcal{E}_{\text{KG}}(q) & t = \text{\code{KG-Fact}}.
    \end{cases}
\]

After retrieval, the system constructs a task-specific prompt $\mathcal{P}$ by concatenating the original user query $q$ with retrieved evidence $\mathcal{E}(q, t)$. The retrieved evidence may consist of text passages or structured knowledge-graph outputs depending on the routing decision. 
\[
    \mathcal{P}(q, \mathcal{E}, t) = \text{Template}(t) \; \Vert \; \mathcal{E}(q, t) \; \Vert \; q,
\]
where $\Vert$ denotes string concatenation and $\text{Template}(t)$ is an instruction prefix corresponding to the predicted task (e.g., \textit{``Summarize the following paper:''}, \textit{``Answer the following question...''}). This explicit prompt structure encourages the generator to rely on retrieved evidence rather than parametric memory.

\subsection{Response Generation}
Instead of relying on large proprietary models, we employ a compact open model (Llama 3.2 3B Instruct) as the response generator. 
This design choice reflects our central hypothesis: strong retrieval can compensate for smaller model capacity. 
The model is instruction-tuned on the OpenScholar~\cite{asaiOpenScholarSynthesizingScientific2024} dataset to support scholarly tasks including multi-paper QA, summarization, and editing.

Fine-tuning is performed using the \code{SFTTrainer} framework \cite{vonwerra2020trl} with BF16 precision. 
Multiple variants are trained with either Low-Rank Adapters (LoRA) \cite{hu2022lora} of rank $r=32$ or full fine-tuning. Training configurations include variable numbers of epochs $\{1, 2, 5\}$ and context lengths $\{8192, 10000, 16384\}$. All models use the \code{AdamW} optimizer \cite{loshchilov2018decoupled} with a \textit{cosine-annealing} learning rate schedule, initial learning rate $5\cdot10^{-6}$, 200 warmup steps, and an effective batch size of 64. We provide the detailed implementation settings in our repository.

During inference, the system receives the composed prompt
$\mathcal{P}(q, \mathcal{E}, t)$ constructed in the previous stage, the model generates a response $r$ and a set of citations $\mathbf{C}$:

\[
\begin{aligned}
r, \mathbf{C} &= g_\phi(\mathcal{P}(q, \mathcal{E}, t)) \\
g_\phi &: \mathcal{P} \mapsto (\text{Textual Answer}, \text{Citations})
\end{aligned}
\]

where $g_\phi$ denotes the fine-tuned LLM parameterized by $\phi$. The generated answer includes references corresponding to the retrieved evidence passages. Model serving is performed using \code{vLLM} \cite{kwon2023efficientvllm}.

\begin{table*}[t]
\centering
\footnotesize
\begin{tabular}{lrrr|r|r|ccc}
\toprule
\textbf{Model} & \textbf{Org$\uparrow$} & \textbf{Cov$\uparrow$} & \textbf{Rel$\uparrow$} & \textbf{Mean$\uparrow$} & \textbf{Length} & \multicolumn{3}{c}{\textbf{Citation Quality}} \\
\cmidrule(lr){7-9}
 &  &  &  &  &  & \textbf{R$\uparrow$} & \textbf{P$\uparrow$} & \textbf{F1$\uparrow$} \\
\midrule

Llama3-8B $\dagger$* & -- & -- & -- & 3.79 & -- & -- & -- & 0.0\% \\
OS-8B * & 3.92 & 4.44 & 4.02 & 4.12 & 578.6 & -- & -- & 42.8\% \\

\midrule

Llama 3.2 3B Instruct $\dagger$ & 3.96 & 4.06 & 4.54 & 4.19 & 450 & 0.00\% & 0.00\% & 0.00\% \\
Llama 3.1 8B Instruct $\dagger$ & 4.01 & 4.19 & 4.76 & 4.32 & 475 & 0.00\% & 0.00\% & 0.00\% \\

\midrule

Llama 3.2 3B Instruct & 3.67 & 3.51 & 3.99 & 3.72 & 363 & 16.49\% & 19.05\% & 17.68\% \\
Llama 3.1 8B Instruct & \underline{3.83} & \underline{3.90} & \underline{4.38} & \underline{\textbf{4.04}} & 453 & \underline{23.50\%} & \underline{\textbf{26.45\%}} & \underline{\textbf{24.89\%}} \\

\midrule

LoRA FT 1ep 8k & \underline{3.74} & 3.78 & 4.20 & 3.91 & 784 & 23.41\% & 21.06\% & 22.18\% \\
LoRA FT 1ep 16k & 3.60 & 3.87 & 4.08 & 3.85 & 1,017 & 23.35\% & 21.63\% & 22.46\% \\
LoRA FT 2ep 8k & 3.65 & 3.81 & 4.18 & 3.88 & 841 & 24.74\% & 23.72\% & 24.22\% \\
LoRA FT 2ep 10k & 3.59 & \underline{\textbf{3.92}} & \underline{\textbf{4.31}} & \underline{3.94} & 1,297 & \underline{\textbf{25.64\%}} & 22.77\% & 24.12\% \\
LoRA FT 2ep 16k & 3.55 & 3.82 & 4.21 & 3.86 & 887 & 24.07\% & \underline{25.02\%} & \underline{24.53\%} \\
LoRA FT 5ep 16k & 3.55 & 3.80 & 4.27 & 3.87 & 1,024 & 20.10\% & 18.75\% & 19.40\% \\

\midrule

Full FT 1ep 16k & 3.81 & 3.80 & 4.20 & 3.94 & 532 & 24.51\% & 23.33\% & \underline{23.91\%} \\
Full FT 2ep 10k & 3.85 & 3.83 & 4.19 & 3.96 & 468 & \underline{24.52\%} & 22.69\% & 23.57\% \\
Full FT 2ep 16k & \underline{\textbf{3.89}} & \underline{3.88} & 4.23 & \underline{4.00} & 466 & 21.20\% & 19.66\% & 20.40\% \\
Full FT 5ep 16k & 3.82 & 3.84 & \underline{4.26} & 3.98 & 484 & 23.81\% & \underline{23.96\%} & 23.89\% \\

\bottomrule
\end{tabular}
\caption{Multi-Paper QA Evaluation results (ScholarQABench-Multi dataset). Best value per \textit{model type} is \underline{underlined}; overall best values (excluding external baselines) are \textbf{bolded}; LoRA FT rows denote LoRA fine-tuning applied to Llama 3.1 8B Instruct; * results from \cite{asaiOpenScholarSynthesizingScientific2024};  
$\dagger$ without retrieval. 
}
\label{tab:scholarqabench_multi_qa}
\end{table*}

\section{Evaluation}
We evaluate our framework to assess whether our design combining task-aware retrieval and lightweight model can effectively support scientific applications. Our evaluation focuses on the extent to which small language models, when combined with task-aware retrieval strategies, can achieve competitive performance on scholarly tasks. In addition, we examine the robustness of the system under varying retrieval conditions, including differences in retrieval quality and domain alignment between the corpus and evaluation datasets.  We select evaluation datasets to reflect complementary aspects of scholarly tasks:
\begin{enumerate}
    \item \textbf{ScholarQABench-Multi~\citeplanguageresource{openscholar}} for multi-document QA and reasoning,
    \item \textbf{PubMedQA~\citeplanguageresource{pubmedqa}} for domain transfer and robustness, and
    \item \textbf{SciTLDR~\citeplanguageresource{tldr}} for extreme summarization.
\end{enumerate}

All experiments follow the pipeline described in Section~\ref{sec:pipeline}. Incoming queries are first routed to a task category, relevant context is retrieved, and the composed prompt is passed to the language model for response generation. 
We evaluate the framework on two primary scholarly tasks: \textbf{scientific question answering} (\S\ref{sec:qa_evaluation}) and \textbf{text compression} (\S\ref{sec:compress_eval}), detailed in the following sections.

\begin{table*}[t]
    \centering
    \footnotesize
    \begin{tabular}{lcccccc}
        \toprule
        \textbf{Model} &
        \makecell{Orig\\Acc$\uparrow$} & \makecell{Orig\\F1$\uparrow$} &
        \makecell{OS\\Acc$\uparrow$} & \makecell{OS\\F1$\uparrow$} &
        \makecell{Zero\\Acc$\uparrow$} & \makecell{Zero\\F1$\uparrow$} \\
        \midrule
        BioBERT $\dagger$           & 68.08 & 52.72 & --    & --    & --    & --    \\
        OS-8B *                     & --    & --    & \textbf{76.4}  & --    & --    & --    \\
        \midrule
        Llama 3.2 3B Instruct       & 62.40 & 40.93 & 49.11 & 49.04 & 64.41 & 58.97 \\
        Llama 3.1 8B Instruct       & \textbf{75.60} & \textbf{54.06} & 45.91 & 43.35 & \textbf{64.77} & \textbf{59.74} \\
        \midrule
        LoRA FT 2ep 10k context     & 68.00 & 46.69 & 58.01 & 55.83 & 58.13 & 57.23 \\
        Full FT 5ep 16k context     & 62.80 & 38.69 & 58.01 & \textbf{56.21} & 60.62 & 57.24 \\
        \bottomrule
    \end{tabular}
    \caption{
        Single-paper QA evaluation results (PubMedQA dataset). Best values are \textbf{bolded}. $\dagger$ results from \cite{jin_pubmedqa_2019}; * results from \cite{asaiOpenScholarSynthesizingScientific2024}.
    }
    \label{tab:pubmedqa}
\end{table*}

\begin{table*}[t]
    \centering
    \footnotesize
    \vspace{3mm}
    \begin{tabular}{lccc|c|c|c}
        \toprule
        \textbf{Model} &
        \makecell{R1\\F1$\uparrow$} &
        \makecell{R2\\F1$\uparrow$} &
        \makecell{RL\\F1$\uparrow$} &
        \makecell{BERTScore\\F1$\uparrow$} &
        \makecell{SMOG\\Index$\uparrow$} &
        \makecell{Compression\\Ratio$\uparrow$} \\
        \midrule
        CATT~\cite{cachola-etal-2020-tldr} & \textbf{44.9}  & \textbf{22.6}  & \textbf{37.3}  & --    & --    & \textbf{47.3} \\
        \midrule
        Llama 3.2 3B Instruct              & 1.21  & 0.011 & 1.21  & 54.55 & 24.45 & 20.72 \\
        Llama 3.1 8B Instruct              & 1.21  & \textbf{0.014} & 1.21  & 54.57 & \textbf{24.81} & 20.79 \\
        \midrule
        Full FT 5ep 16k context            & \textbf{1.31}  & \textbf{0.014} & \textbf{1.30}  & \textbf{54.74} & 23.26 & \textbf{22.43} \\
        \bottomrule
    \end{tabular}
    \caption{
        Text compression evaluation results (SciTLDR dataset). Best values are \textbf{bolded}. R1, R2, and RL denote ROUGE-1, ROUGE-2, and ROUGE-L F1 scores. Scores are averages over test samples.
    }
    \label{tab:SciTLDR}
\end{table*}

\subsection{Scientific Question Answering Evaluation}
\label{sec:qa_evaluation}
For scientific question answering, we evaluate in two settings: (1) multi-paper QA (\S\ref{sec:multi}), which requires synthesizing information across several documents, and (2) single-paper QA (\S\ref{sec:single}), focusing on questions grounded in a single paper.

\subsubsection{Multi-Paper Question Answering}
\label{sec:multi}
We evaluate the framework on multi-paper question answering using the ScholarQABench-Multi benchmark~\citeplanguageresource{openscholar}. The dataset contains 108 questions spanning computer science, physics, and biomedical research. Answers are evaluated using LLM-based judges (Prometheus models) across three dimensions: Organization (\textit{Org}), Relevance (\textit{Rel}), and Coverage (\textit{Cov}). Each dimension is scored on a 1–5 scale and averaged to obtain an overall quality score (\textit{Mean}). In addition, citation quality is measured using Precision, Recall, and F1 with respect to gold references. Table~\ref{tab:scholarqabench_multi_qa} summarizes our evaluation results.

\paragraph{Impact of Retrieval.}  
We first compare base models with and without retrieval augmentation. Surprisingly, LLM evaluation scores are slightly higher when retrieval is disabled. This suggests that long prompts containing multiple retrieved passages can make generation more difficult for smaller models. However, without retrieval the models are unable to produce verifiable citations as expected, resulting in a citation F1 score of 0. Retrieval therefore remains essential for grounding responses in evidence.

\paragraph{Fine-Tuning Effects.}  
Fine-tuning substantially improves both answer quality and citation grounding. Our fully fine-tuned 3B model approaches the performance of a substantially larger 8B model. Low-Rank Adaptation (LoRA) improves performance relative to the base model but occasionally produces repetitive outputs or unstable citation formatting, which negatively affects organization scores. Across experiments, training duration and context length have limited impact on overall performance, though longer training slightly degrades organization for LoRA models.

These results on multi-paper QA highlight a key trade-off: while retrieval enables citation grounding, it can introduce noise that negatively affects answer organization and fluency, particularly for smaller models with limited context handling capacity.

\subsubsection{Single-Paper Question Answering}
\label{sec:single}
We evaluate single-paper question answering using the biomedical PubMedQA~\citeplanguageresource{pubmedqa} dataset to further assess retrieval robustness and domain generalization. 
The task requires predicting \textit{yes}, \textit{maybe}, or \textit{no} answers to biomedical questions. We report our evaluation results for the best-performing LoRA and full fine-tuned models, as well as pre-trained Llama 3.1 8B and 3B models as comparison. The results are summarized in Table~\ref{tab:pubmedqa}. 

We follow the retrieval-based evaluation protocol introduced in ScholarQABench~\citeplanguageresource{openscholar} and consider the following three evaluation settings.

\vspace{-3mm}
\paragraph{Original Task (Orig).} 
In the original task setup, models are provided with gold context passages from the relevant paper abstracts. Under this setup, larger models achieve higher accuracy, indicating stronger reasoning ability when reliable evidence is available. The LoRA-adapted 3B model shows notable improvement over its base version despite not being trained on PubMedQA, suggesting that lightweight adaptation can improve domain transfer.

\paragraph{Retrieval Task (OS)}

In the retrieval variant, models must identify relevant passages from the corpus. The performance drop in the retrieval setting reflects a domain mismatch between the unarXive corpus (primarily computer science and physics) and PubMedQA (biomedical domain). This setup allows us to explicitly evaluate robustness under domain shift.
Nevertheless, we show that the fine-tuned models outperform their pretrained counterparts, indicating that training improves the model’s ability to filter irrelevant context by ignoring irrelevant retrievals and leveraging retrieved content.

\vspace{-3mm}
\paragraph{Zero-Context Task (Zero).}
Finally, we evaluate models without any retrieved evidence. In this setting, answers therefore rely entirely on the models’ parametric knowledge. Pretrained models perform slightly better than fine-tuned ones, suggesting that task-specific training may introduce mild specialization that reduces general-domain recall. This highlights the trade-off between fine-tuning for domain-specific QA and preserving general domain knowledge.

\subsection{Text Compression Evaluation}
\label{sec:compress_eval}
To evaluate summarization capabilities, we benchmark the system on the SciTLDR~\citeplanguageresource{tldr} dataset, which focuses on extreme compression of scientific papers. The task requires generating a one-sentence summary from the abstract, introduction, and conclusion (AIC) sections. Table~\ref{tab:SciTLDR} summarizes our evaluation results.

Generated summaries are compared against the gold TLDR statement summaries reviewed by authors and peer reviewers. 
Overlap metrics were computed using ROUGE-1, ROUGE-2, and ROUGE-L \cite{lin-2004-rouge}, as well as BERTScore \cite{bert-score}.
Following prior work~\cite{cachola-etal-2020-tldr}, we report the maximum target score for each metric to reduce variability across reference summaries. In addition, we compute a compression ratio between the source text and generated summaries. Because individual summaries consist of a single sentence, traditional readability metrics (\textit{Flesch, Flesch-Kincaid, SMOG}) cannot be applied per example. Thus, we concatenate outputs and compute aggregate readability scores\footnote{\url{https://github.com/cdimascio/py-readability-metrics}} to approximate lexical and syntactic complexity.

We compare our fine-tuned 3B model with baseline Llama models and the specialized CATT model used in the original SciTLDR benchmark. The results show modest improvements from fine-tuning. ROUGE and BERTScore increase slightly, and the compression ratio improves relative to the base model. However, generated summaries remain longer than the target TLDR statements, and readability metrics suggest relatively dense language. These findings highlight the difficulty of extreme scientific summarization and confirm that models specifically trained for the task continue to outperform general-purpose LLMs.

\section{Conclusion}
\label{sec:conclusion}

We presented a lightweight retrieval-augmented framework for scholarly assistance that combines task-aware routing, hybrid retrieval, and compact language models within a unified architecture. Rather than relying on increasingly large proprietary systems, our work investigates under which conditions improved retrieval design can compensate for reduced model scale across scholarly tasks.

Our evaluation results show that small instruction-tuned models can approach the performance of larger systems when paired with appropriate retrieval strategies, particularly for tasks requiring grounded, citation-based answers. Fine-tuning further improves robustness in multi-document question answering and helps mitigate the impact of partially irrelevant context. However, these gains are not uniform: model capacity remains a key factor for reasoning quality, especially in complex or domain-shifted settings.

Importantly, our findings highlight that improvements in retrieval introduce inherent trade-offs. While more precise retrieval can improve grounding and interpretability, it may reduce recall or introduce longer, noisier contexts that disproportionately affect smaller models. In addition, retrieval pipelines incur additional engineering complexity and may generalize poorly when the underlying corpus does not match the target domain, as observed in our biomedical QA evaluation. These results emphasize that retrieval and model scale should be viewed as complementary components rather than interchangeable solutions.

Overall, our study suggests that progress in retrieval quality is as critical as progress in model scaling for building practical and efficient scholarly assistants. 
Future work will focus on improving retrieval robustness, expanding domain coverage, and incorporating automated verification of generated outputs. We also plan to explore efficiency-oriented improvements, such as more accurate routing strategies and lightweight reranking, alongside broader evaluations that assess usability, readability, and trust in real-world scholarly workflows.

\section{Limitations}  

First, although the proposed design integrates structured scholarly metadata from knowledge graph and textual evidence within a single pipeline, a standardized benchmark for question answering over the SemOpenAlex knowledge graph is currently unavailable. Thus, current evaluation focuses on the text-based components, while the KG-Fact module is presented primarily as an architectural capability. Constructing reliable SPARQL-question pairs for large scholarly knowledge graphs remains an open challenge. Second, retrieval quality remains a bottleneck. Dense vector search may return passages that are only loosely related to the query, reducing grounding quality and potentially leading to weak or partially supported citations. 

Finally, the current system does not include a post-generation verification step to validate citations against source documents. Incorporating reference validation or retrieval-based fact checking would be an important step toward more reliable scholarly assistants.

\section{Ethical Considerations}  
All datasets used are publicly available under research-friendly licenses, e.g., ScholarQABench-Multi~\citeplanguageresource{openscholar} is released under the ODC-BY license, with some constituent datasets subject to their own licensing terms. Our 165K-paper datastore unarXive~\citeplanguageresource{unarXive} comprises open-access content compliant with text and data mining permissions. The system is designed as an assistive tool for scholarly tasks and is not intended for use in high-stakes domains without human oversight. We acknowledge potential biases in the underlying literature (e.g., publication bias, domain skew) that may influence system behavior. Additionally, our approach encourages the use of smaller or lightweight models (e.g., 3B parameters) over larger models to reduce computational and environmental impact.

\section{Bibliographical References}\label{sec:reference}

\bibliographystyle{lrec2026-natbib}
\bibliography{lrec2026-example,ref}

\section{Language Resource References}
\label{lr:ref}
\bibliographystylelanguageresource{lrec2026-natbib}
\bibliographylanguageresource{languageresource}

\end{document}